\documentstyle[aps,12pt]{revtex}

\begin{document}
\author{Z.-D. Chen$^{1,2,3}$, J.-Q. Liang$^1$, S.-Q. Shen$^4$, W.-F. Xie$^2$}
\address{1. Institute of Theoretical Physics and Department of Physics, Shanxi\\
University, Taiyuan, Shanxi, 030006, China\\
2. Department of Physics, Guangzhou University, Guangzhou, Guangdung,\\
510405, \ China\\
3. Department of Physics, Jinan University, Guangzhou, Guangdung, 510632,\\
China\\
4. Department of Physics, The University of Hong Kong, Hong Kong, China}
\title{Dynamics and Berry Phase of Two-Species Bose-Einstein Condensates}
\maketitle

\begin{abstract}
In terms of exact solutions of the time-dependent Schr$\ddot{o}$dinger
equation for an effective giant spin modeled from a coupled two-mode
Bose-Einstein condensate (BEC) with adiabatic and cyclic time-varying Raman
coupling between two hyperfine states of the BEC we obtain analytic
time-evolution formulas of the population imbalance and relative phase
between two components with various initial states especially the SU(2)
coherent state. We find the Berry phase depending on the number parity of
atoms, and particle number dependence of the collapse-revival of
population-imbalance oscillation. It is shown that self-trapping and phase
locking can be achieved from initial SU(2) coherent states with proper
parameters.

PACS numbers, 03.75. Kk, 05.30.Jp, 32.80.Pj, 03.65.Fd
\end{abstract}

\section{Introduction}

The experimental discovery of Bose-Einstein condensation in trapped atomic
clouds opened up the exploration of quantum mechanics of mesoscopic systems
in qualitatively new regime. The cold gas clouds have many advantages for
investigations of quantum phenomena and hence become a test ground of
quantum mechanical principles as well as the interplay between macroscopics
and quantum coherence. The observation of matter-wave interference implies
the realization of coherent atomic beams, atomic Josephson effect and a
variety of quantum interference phenomena\cite{1}. In particular, recent
experiments on two-component Bose-Einstein condensates (BECs) in $^{87}$Rb
atoms\cite{1,2} have stimulated considerable interests in the phase dynamics
and number fluctuations of the condensates. Aside from its intrinsic appeal,
the capability demonstrated by the recent experiments might lead to
applications also on quantum computation. Based on the macroscopic wave
function approach it is demonstrated that the Josephson effect exists in a
driven two-state single-particle BEC in a single trap. The macroscopic
quantum self-trapping as well as the $\pi $-phase oscillations in which the
time averaged value of the phase difference is equal to $\pi $ have been
studied extensively\cite{3,4,5,6,7}. It is also shown that the population
oscillation is modulated by the collapses and revivals due to the quantum
nature of the system\cite{8,9,10,11,12,13,14}. The relative phase of two
condensates in different hyperfine atomic states can be measured\cite{2}
using Ramsey's method of separated oscillating fields\cite{15} and it is
evident that the phase locking indeed occurs for small separation between
condensates\cite{1}, implying the broken gauge symmetry. Most theoretical
studies are focused on semiclassical analysis and a full quantum mechanical
formulation of the dynamics of the two-component BECs coupled by
time-dependent driving is certainly of interest and importance. It is well
known that the system of two-component BECs can be described by a giant or
mesoscopic pseudo-spin\cite{16} using Schwinger realization of angular
momentum operators in terms of two-mode bosons. Berry phase\cite{17} emerges
naturally in the mesoscopic pseudo-spin model\cite{16} if the coupling
between two components varies with time cyclically and adiabatically. The
Berry phase in the mesoscopic spin model for the coupled two-component BECs
has been explored recently in an elegant way by means of geometric evolutions%
\cite{16}. Trapped atomic BECs make it possible to create mesoscopic quantum
objects containing of the order 10$^6$ atoms in the same quantum state with
a longer life time allowing the implementation of adiabatic evolution which
is required for the Berry Phase. However, the dynamics of the mesoscopic
spin modeled from the two-mode BEC has not yet been studied in the quantum
mechanical formalism. We in the present paper use the exact solution of the
time-dependent Schr$\ddot{o}$dinger equation for the mesoscopic spin to
provide a quantum mechanical evaluation of the phase dynamics and the number
fluctuation. With the time-evolution operator obtained by means of the
generator of time-dependent SU(2) coherent states\cite{18} we are able to
derive analytic time-evolution formulas of both the population imbalance and
relative phase between the two-component BECs for various initial states, in
particular, the SU(2) coherent state with which the new effect of particle
number dependence is discovered. Moreover, our approach has advantage to
obtain the phase dynamics and number fluctuation for both cases with and
without the nonlinear interatomic collisions in the same framework so that
the effects of interatomic collision can be recognized explicitly by
comparing the results between two cases. We show that interatomic collisions
do not affect the Berry phase but lead to the damping and collapse-revival
of the population-imbalance oscillation depending explicitly on the coupling
strength and the total number of atoms as well. The SU(2) coherent states
are the most realistic initial states for the two-species BEC created by
coupling two hyperfine states of \thinspace atoms with radiation field\cite
{12}. To our knowledge we in this paper report, for the first time, a full
quantum mechanical evaluation of the dynamics of the two-species BEC
described by an explicitly time-dependent Hamiltonian with the initial SU(2)
coherent states and the novel phenomena such as self-trapping, phase
locking, and collapse-revival are recovered theoretically in the same
formalism.

The plan of the paper is as follows. In Sec. II we derive the mesoscopic
spin model from the two-component BECs. The SU(2) coherent states, which are
considered as most practical initial states for the two-species BEC, are
briefly introduced. The dynamics and Berry phase of the pseudo-spin are
investigated in terms of the SU(2) coherent state technique for both cases
with and without the interatomic collisions in Sec. III.

\section{Model and Initial state}

We consider Bose-Einstein condensate of trapped atomic gas in a single trap
consisting of two internal states which are coupled by a spatially uniform
radiation field with a Rabi frequency $\Lambda $. Atoms are subjected to
trapping potential V$_l$ (l=a,b). The atoms interact via elastic two-body
collisions with the interaction potential of $\delta $-function type. In the
formalism of second quantization, the system is described by the Hamilton
operator 
\begin{equation}
\widehat{H}=\sum_{l=a,b}\widehat{H}_l+\widehat{H}_{int}+\widehat{H}_f
\end{equation}
\begin{equation}
\widehat{H}_l=\int d^3{\bf r}\left\{ \widehat{\Psi }_l^{+}({\bf r})\left[ -%
\frac{\hbar ^2}{2m}\bigtriangledown ^2+V_l({\bf r})\right] \widehat{\Psi }_l(%
{\bf r})+\frac{q_l}2\widehat{\Psi }_l^{+}({\bf r})\widehat{\Psi }_l^{+}({\bf %
r})\widehat{\Psi }_l({\bf r})\widehat{\Psi }_l({\bf r})\right\}
\end{equation}
\begin{equation}
\widehat{H}_{int}=q_{a,b}\int d^3{\bf r}\widehat{\Psi }_a^{+}({\bf r})%
\widehat{\Psi }_b^{+}({\bf r})\widehat{\Psi }_a({\bf r})\widehat{\Psi }_b(%
{\bf r})
\end{equation}
\begin{equation}
\widehat{H}_f=\Lambda (t)\int d^3{\bf r}\left( \widehat{\Psi }_a^{+}({\bf r})%
\widehat{\Psi }_b({\bf r})e^{i\varphi (t)}+\widehat{\Psi }_b^{+}({\bf r})%
\widehat{\Psi }_a({\bf r})e^{-i\varphi (t)}\right)
\end{equation}
We here have used the field interaction representation in rotating frame. \
The Rabi frequency $\Lambda (t)$ is time-dependent in the sense that it can
be turned on and off adiabatically\cite{3}. The phase $\varphi (t)$ due to
the small detuning of external field from resonance excitation varies with
time slowly and therefore we work on the adiabatically time-varying
Hamiltonian. The phase $\varphi (t)$ which we will see plays a central role
in generating of the Berry phase.

\bigskip In the two-mode approximation of condensation such that $\widehat{%
\Psi }_a({\bf r})\approx \widehat{a}\phi _a({\bf r}),\widehat{\Psi }_b({\bf r%
})\approx \widehat{b}\phi _b({\bf r})$ where $\widehat{a},$ $\widehat{b}$
are the annihilation operators obeying the usual boson commutation
relations, we have

\begin{equation}
\widehat{H}_a=\omega _a\widehat{a}^{+}\widehat{a}+\frac{\eta _a}2\widehat{a}%
^{+}\widehat{a}^{+}\widehat{a}\widehat{a}
\end{equation}
\begin{equation}
\widehat{H}_b=\omega _b\widehat{b}^{+}\widehat{b}+\frac{\eta _b}2\widehat{b}%
^{+}\widehat{b}^{+}\widehat{b}\widehat{b}
\end{equation}
with 
\begin{equation}
\omega _l=\int d^3{\bf r}\phi _l^{*}({\bf r})\left[ -\frac{\hbar ^2}{2m}%
\bigtriangledown ^2+V_l({\bf r})\right] \phi _l({\bf r}),\qquad l=a,b\quad ,
\end{equation}
and 
\[
\eta _l=q_l\int d^3{\bf r|}\phi _l({\bf r})|^4,\quad l=a,b\quad . 
\]
The interaction operator between two species of atoms is 
\begin{equation}
\widehat{H}_{int}=\chi \widehat{a}^{+}\widehat{a}\widehat{b}^{+}\widehat{b}%
\quad ,
\end{equation}
where 
\[
\chi =q_{a,b}\int d^3{\bf r|}\phi _a({\bf r})|^2{\bf |}\phi _b({\bf r})|^2. 
\]
The transition operator induced by the external field is 
\begin{equation}
\widehat{H}_f=G(t)(\widehat{a}^{+}\widehat{b}e^{i\varphi (t)}+\widehat{b}^{+}%
\widehat{a}e^{-i\varphi (t)})
\end{equation}
with 
\[
G(t)=\Lambda (t)\int d^3{\bf r}\phi _a^{*}({\bf r})\phi _b({\bf r}). 
\]
The time-dependent coupling drive is characterized by its Rabi frequency $%
\Lambda (t)$ and the phase $\varphi (t)$. Here we consider the adiabatically
varying phase, \ $\varphi (t)$ ,such that its time-derivative is negligibly
small. The study of dynamics of the system would be greatly simplified by
introducing of the pseudo-angular momentum operators in terms of Schwinger
relation, 
\begin{equation}
\widehat{J}_x=\frac 12(\widehat{a}^{+}\widehat{b}+\widehat{b}^{+}\widehat{a})
\end{equation}
\begin{equation}
\widehat{J}_y=\frac 1{2i}(\widehat{a}^{+}\widehat{b}-\widehat{b}^{+}\widehat{%
a})
\end{equation}
\begin{equation}
\widehat{J}_z=\frac 12(\widehat{a}^{+}\widehat{a}-\widehat{b}^{+}\widehat{b})
\end{equation}
The Casimir invariant is 
\begin{equation}
\widehat{J^2}=\frac{\widehat{N}}2(\frac{\widehat{N}}2+1)
\end{equation}
where $\widehat{N}=\widehat{a}^{+}\widehat{a}+\widehat{b}^{+}\widehat{b}$ is
the total number operator, which is a conserved quantity and thus is set
equal to the total number of atoms N=2j with j being the quantum number of
angular momentum. The Hamilton operator apart from a trivial constant reads 
\begin{equation}
\widehat{H}=\omega _0\widehat{J}_z+q\widehat{J}_z^2+G(t)(\widehat{J}%
_{+}e^{i\varphi (t)}+\widehat{J}_{-}e^{-i\varphi (t)})
\end{equation}
where $\omega _0=\omega _a-\omega _b+(N-1)\frac{\eta _a-\eta _b}2$, $q$=$%
\frac{\eta _a+\eta _b}2-\chi $ , and $\widehat{J}_{\pm }=\widehat{J}_x\pm i%
\widehat{J}_y.$

The relative phase of two-mode BEC surely can be abstracted from the
expectation value of the angular momentum operators $\widehat{J}_{\pm }$
which along with the expectation value of $\widehat{J}_z$ (giving rise to
the population imbalance between two components of BECs) is measurable
quantity in experiment.

Since the two-component BECs are experimentally created by coupling two
hyperfine states with radiation field, it has been shown that the prepared
initial state can be a particular case of the SU(2) coherent state\cite{12}
(or known as atomic coherent state in quantum optics) which describes a
state with a well defined relative phase between the two species. However, a
full quantum evaluation of the dynamics for the two-species BECs coupled by
time-dependent driving with the initial spin coherent states has not yet
been given. SU(2) coherent state is defined as 
\begin{equation}
\widehat{J}\cdot {\bf n}|{\bf n}\rangle =j|{\bf n}\rangle 
\end{equation}
where ${\bf n}=$($\sin \theta \cos \phi $, $\sin \theta \sin \phi $, $\cos
\theta $) is a unit vector. The SU(2) coherent state can be generated from
an extreme Dicke state such that 
\begin{equation}
\widehat{\Omega }(\theta ,\phi )|j,j\rangle =|{\bf n}\rangle 
\end{equation}
where 
\begin{equation}
\widehat{\Omega }(\theta ,\phi )=e^{\frac \theta 2(\widehat{J}_{-}e^{i\phi }-%
\widehat{J}_{+}e^{-i\phi })}
\end{equation}
This is called the coherent state in the north pole gauge, as compared to
the generation of the coherent state from the extreme state $|j,-j\rangle $
where it is called the state in the south pole gauge. The Dicke states are
defined as usual $\widehat{J}_z|j,m\rangle =m|j,m\rangle $ and can be
generated from the vacuum by boson creation operators, i.e. 
\begin{equation}
|j,m\rangle =\frac 1{\sqrt{(j+m)!(j-m)!}}(\widehat{a}^{+})^{j+m}(\widehat{b}%
^{+})^{j-m}|0\rangle 
\end{equation}
The coherent state of eq.(16) can be expanded in terms of the Dicke states, 
\begin{equation}
|{\bf n}\rangle =\sum_{m=-j}^j%
{2j \choose j+m}%
^{\frac 12}(\cos \frac \theta 2)^{j+m}(\sin \frac \theta 2%
)^{j-m}e^{i(j-m)\phi }|j,m\rangle 
\end{equation}
It is easy to verify that 
\begin{equation}
\langle \widehat{J}_z\rangle =\langle {\bf n|}\widehat{J}_z|{\bf n}\rangle =%
\frac N2\cos \theta ,\quad \langle \widehat{J}_{+}\rangle =\frac N2\sin
\theta e^{i\phi },\quad \langle \widehat{J}_{-}\rangle =\frac N2\sin \theta
e^{-i\phi }
\end{equation}
and the phase $e^{i\phi }$ is seen to be the relative phase of the two
species prepared in the initial SU(2) coherent state. In this paper the
generator of SU(2) coherent states eq.(17) is used as a unitary
transformation to formulate the dynamics of the mesoscopic spin system
modeled from the two-component BECs.

\bigskip

\bigskip

\section{Dynamics and Berry Phase}

\subsection{ The case of q=0}

We first of all consider the case of $q=0$ achieved by the condition, $\frac{%
\eta _a+\eta _b}2=\chi $ , which, although a special case, is practical in
the range of BEC\ parameters. The equality such that $q_a\approx q_b\approx $
$q_{ab}$ can be fulfilled practically for the two-component BECs consisting
of $^{87}R_b$ atoms with different internal states since the scattering
lengths of atoms with the two internal states are known at the $1\%$ level
to be\cite{2} in proportion $a_a:a_{ab}:a_b=1.03:1:0.97$ where $a_a$ and $a_b
$ are the same-species scattering lenghts and $a_{ab}$ is the scattering
length for interspecies collisions. The ground state wave function is same, $%
\phi _a({\bf r})=\phi _b({\bf r})$, and therefore the condition $\eta
_a\approx \eta _b\approx \chi $ can be satisfied. The model in the case of $%
q=0$ is exactly solvable. We start with a generalized gauge transformation%
\cite{19} in terms of the time-dependent unitary transformation\cite{20}
defined by 
\begin{equation}
\widehat{R}(t)=e^{\frac \lambda 2(\widehat{J}_{-}e^{-i\varphi (t)}-\widehat{J%
}_{+}e^{i\varphi (t)})}
\end{equation}
which has the same form as the generator eq.(17) of SU(2) coherent states
and is the key point of the present formulation. The time-dependent
Schr\"{o}dinger equation is convariant under the gauge transformation\cite
{19} such that 
\begin{equation}
i\frac d{dt}|\psi ^{\prime }(t)\rangle =\widehat{H}^{\prime }|\psi ^{\prime
}(t)\rangle 
\end{equation}
where 
\begin{equation}
\widehat{H}^{\prime }=\widehat{R}\widehat{H}\widehat{R}^{+}-i\widehat{R}%
\frac \partial {\partial t}\widehat{R}^{+},\qquad |\psi ^{\prime }(t)\rangle
=\widehat{R}|\psi (t)\rangle 
\end{equation}
and the state $|\psi (t)\rangle $ is assumed to be the solution of original
Schr\"{o}dinger equation 
\begin{equation}
i\frac d{dt}|\psi (t)\rangle =\widehat{H}|\psi (t)\rangle 
\end{equation}
The auxiliary parameter $\lambda $ ,which is time-dependent in general, is
to be determined by requiring that the Hamilton operator $\widehat{H}%
^{\prime }$ is diagonal in the $\widehat{J}_z$ representation. Using the
relations given in appendix (eqs.A1-A4)\cite{20}, and noticing the adiabatic
condition that $\frac{d\varphi }{dt}\simeq 0$ and $\frac{d\lambda }{dt}%
\simeq 0$, we obtain the Hamilton operator 
\begin{equation}
\widehat{H}^{^{\prime }}=\alpha (t)\widehat{J}_z\qquad \qquad \alpha (t)=%
\sqrt{\omega _0^2+4G^2(t)}
\end{equation}
with auxiliary parameter $\lambda $ chosen as 
\begin{equation}
\sin \lambda =-\frac{2G(t)}{\omega _0}\cos \lambda ,
\end{equation}
and
\begin{equation}
\qquad \cos \lambda =\frac{\omega _0}{\alpha (t)}.
\end{equation}
It is easy to obtain the exact general solution of the original
Schr\"{o}dinger equation 
\begin{equation}
|\psi (t)\rangle =\sum_{m=-j}^jc_me^{-i\alpha _m(t)}|j,m(t)\rangle ,\qquad
|j,m(t)\rangle =\widehat{R}^{+}(t)|j,m\rangle 
\end{equation}
with $|j,m\rangle $ being the usual eigenstate of angular momentum $\widehat{%
J}_z$ that $\widehat{J}_z|j,m\rangle =m|j,m\rangle $. The total phase is
seen to be 
\begin{equation}
\alpha _m(t)=\varepsilon _m(t)+\gamma _m(t)=m\int_0^t\alpha (t^{^{\prime
}})dt^{^{\prime }}
\end{equation}
which consists of the dynamical part given by 
\begin{equation}
\varepsilon _m(t)=\int_0^t\langle j,m(t^{\prime })|\widehat{H}|j,m(t^{\prime
})\rangle dt^{\prime }
\end{equation}
and the geometric part i.e. the Berry phase 
\begin{equation}
\gamma _m(t)=-i\int_0^t\langle j,m(t^{\prime })|\frac \partial {\partial
t^{\prime }}|j,m(t^{\prime })\rangle dt^{\prime }
\end{equation}
which is defined in the usual way. In the following we only consider a
time-independent Rabi frequency $G$ for the sake of simplicity. For a
variation of one period $T$ i.e. $\varphi (T)-\varphi (0)=2\pi $ the Berry
phase is found as 
\begin{equation}
\gamma _m(T)=-m\oint (1-\cos \lambda )d\varphi =m\frac{\omega _0-\alpha }%
\alpha 2\pi 
\end{equation}
which has an obvious, geometric meaning from the viewpoint of differential
geometry that the one form $d\varphi $ is exact but not closed. Where $%
\alpha =\sqrt{\omega _0^2+4G^2}$ which is a time-independent parameter. The
Berry phase does not depend on explicit form of the function $\varphi (t)$
and is simply $m$ times of a solid angle with the polar angle $\lambda $ in
agreement with the recent result reported in Ref.\cite{16} in which the
Berry phase is evaluated in terms of geometric evolutions for the coupled
two-component BECs. We, however, following the original procedure of Berry%
\cite{17} obtain the Berry phase and the exact wave function as well by
solving the time-dependent Schr$\ddot{o}$dinger equation. In our approach
the time-evolution of both population imbalance and relative phase between
two components of BECs can be investigated analytically. The explicit
dependence of Berry phase on the parameters of two-species BECs is also
given with our procedure and the novel properties of the Berry phase can be
explored. The geometric phase is actually the same as obtained in the
context of SU(2) coherent state path integrals\cite{21}. To study the
dynamics of the mesoscopic spin it is useful to derive the explicit
time-evolution operator such that 
\begin{equation}
|\psi (t)\rangle =\widehat{U}(t)|\psi (0)\rangle 
\end{equation}
where the time-evolution operator is found from the exact general solution
of the time-dependent Schr$\ddot{o}$dinger equation $|\psi (t)\rangle $ in
eq.(28) as\cite{20} 
\begin{equation}
\widehat{U}(t)=\widehat{R}^{+}(t)e^{-i\alpha t\widehat{J}_z}\widehat{R}(0)
\end{equation}
For a given initial state $|\psi (0)\rangle $ of the system the population
imbalance between two components can be evaluated by 
\begin{equation}
\Delta N(t)=N_a(t)-N_b(t)=2\langle \psi (0)|\widehat{J}_z(t)|\psi (0)\rangle 
\end{equation}
where the time-dependent angular momentum operator $\widehat{J}_z(t)$ in
Heisenberg picture is given in appendix (eq.(A5)) with the help of
eqs.(A1-A4). For the sake of simplicity we have set the initial phase to
zero i.e. $\varphi (0)=0$. To have the phase dynamics we need also the
time-dependent angular momentum $\widehat{J}_{+}(t)$ or $\widehat{J}_{-}(t)$
in Heisenberg picture and the explicit formula of $\widehat{J}_{+}(t)$ is
shown in eq.(A6). The time-evolution of population imbalance and expectation
value of $\widehat{J}_{+}(t)$ given by 
\begin{equation}
\langle \widehat{J}_{+}\rangle =\langle \psi (0)|\widehat{J}_{+}(t)|\psi
(0)\rangle 
\end{equation}
are evaluated for various initial states as follows.

(1)We consider, first of all, the initial state $|\psi (0)\rangle
_1=|j,m\rangle $ and obtain in terms of eqs.(A5), (A6) 
\begin{equation}
\Delta N_1(t)=\frac{2m}{\alpha ^2}[\omega _0^2+4G^2\cos (\alpha t+\varphi
(t))]
\end{equation}
and 
\begin{equation}
\langle \widehat{J}_{+}\rangle _1=m\sin \lambda [-\cos \lambda e^{-i\varphi
(t)}+\cos ^2\frac \lambda 2e^{i\alpha t}-\sin ^2\frac \lambda 2%
e^{-i[2\varphi (t)+\alpha t]}].
\end{equation}
The population imbalance exhibits a simple oscillation. It is interesting to
see a fact that the state of vanishing imbalance can be achieved for the
case of even number of particles (i.e. $j=\frac N2$ is integer) with the
initial state of $m=0$, but not for the case of odd number of particles where%
$\ j$ is half-integer and the state of $m=0$ does not exist.

(2)For a general SU(2) coherent state $|\psi (0)\rangle _2=|{\bf n}\rangle $
, the population imbalance is found as 
\begin{eqnarray}
\Delta N_2(t) &=&\frac N2\{[\cos ^2\lambda +\sin ^2\lambda \cos (\alpha
t+\varphi (t))]\cos \theta -\sin \lambda \cos \lambda \sin \theta \cos \phi
\\
&&+[\cos ^2\frac \lambda 2\cos (\alpha t+\varphi (t)+\phi )-\sin ^2\frac %
\lambda 2\cos (\alpha t+\varphi (t)-\phi )]\sin \lambda \sin \theta \} 
\nonumber
\end{eqnarray}
The self trapping with non-vanishing population imbalance takes place for
the initial state with $\phi =0$ and $\theta =-$ $\lambda $. The population
imbalance thus reduces to 
\begin{equation}
\Delta N_2(t;\phi =0,\theta =-\lambda )=\frac N2\cos \lambda =\frac N2\frac{%
\omega _0}\alpha
\end{equation}
The asymmetric trap potential i.e. non-vanishing $\omega _0$ is the
necessary condition to achieve the self-trapping. The expectation value of
angular momentum operator $\widehat{J}_{+}(t)$ for the initial SU(2)
coherent state is seen to be 
\begin{eqnarray}
\langle \widehat{J}_{+}\rangle _2 &=&\frac N2\{\sin \lambda [\cos ^2\frac %
\lambda 2e^{i\alpha t}-\sin ^2\frac \lambda 2e^{-i[\alpha t+2\varphi
(t)]}-\cos \lambda e^{-i\varphi (t)}]\cos \theta \\
&&+[\cos ^4\frac \lambda 2e^{i\alpha t}+\sin ^4\frac \lambda 2e^{-i[\alpha
t+2\varphi (t)]}+\frac 12\sin ^2\lambda e^{-i\varphi (t)}]e^{i\phi }\sin
\theta  \nonumber \\
&&+[-\cos ^2\frac \lambda 2\sin ^2\frac \lambda 2(e^{i\alpha t}+e^{-i[\alpha
t+2\varphi (t)]})+\frac 12\sin ^2\lambda e^{-i\varphi (t)}]e^{-i\phi }\sin
\theta \}  \nonumber
\end{eqnarray}
and reduces to 
\begin{equation}
\langle \widehat{J}_{+}\rangle _2(\phi =0,\theta =-\lambda )=N\frac G\alpha
e^{-i\varphi (t)}
\end{equation}
for the initial state with $\phi =0$ and $\theta =-$ $\lambda $ indicating
obviously the phase locking$.$ We see that self-trapping of both population
imbalance and relative phase of the two-component BECs can be obtained
simultaneously from the SU(2) coherent state. For the particular case of
symmetric trap potential $\omega _a=\omega _b$ ( $\omega _0=0$ ) we have $%
\lambda =-\frac \pi 2$ seen from eq.(27), and the Berry phase then reduces
to 
\begin{equation}
\gamma _m(T,\omega _0=0)=m\oint d\varphi =m2\pi .
\end{equation}
The population imbalance is 
\begin{equation}
\Delta N_1(t,\omega _0=0)=m\cos (2Gt+\varphi (t))]
\end{equation}
and 
\begin{equation}
\langle \widehat{J}_{+}\rangle _1(\omega _0=0)=m[\frac 12e^{i2Gt}-\frac 12%
e^{-i[2\varphi (t)+2Gt]}]
\end{equation}
The population imbalance for the initial SU(2) coherent state vanishes seen
obviously from eq.(40) in the case of $\omega _0=0$ ,while 
\begin{equation}
\langle \widehat{J}_{+}\rangle _2(\phi =0,\theta =-\lambda ,\omega _0=0)=%
\frac N2e^{-i\varphi (t)}
\end{equation}
The relative phase of two components is locked exactly to the phase of
external field. The phase locking, which we will see, remains in the case
with interatom collisions i.e. the non-vanishing $q$. It may be worthwhile
to emphasize that the Berry phase of eq.(43) is trivial in the case of even
number of particles ($m$ is integer) while it would lead to an antiperiodic
wave function under $2\pi $ evolution of the phase angle $\varphi $ for the
odd number of particles ($m$ is half-integer in this case) similar to the
spin parity effect in the macroscopic quantum coherence in spin systems\cite
{22,23,24}.

\subsection{ With Nonlinear Interactions}

\bigskip We now consider the general case with non-vanishing but small $q$
for the practical model at hand i.e. the two-species BEC created by coupling
two hyperfine states of $^{87}R_b$ atoms. For a single trap a reasonable
condition is $\omega _0=0$ $(\lambda =-\frac \pi 2)$. With the help of
time-dependent unitary transformation eq.(21) and the relation given by 
\begin{eqnarray}
\widehat{R}\widehat{J}_z^2\widehat{R}^{+} &=&\widehat{J}_z^2\cos ^2\lambda +%
\frac 12[(\widehat{J}_z\widehat{J}_{+}+\widehat{J}_{+}\widehat{J}%
_z)e^{i\varphi (t)}+(\widehat{J}_z\widehat{J}_{-}+\widehat{J}_{-}\widehat{J}%
_z)e^{-i\varphi (t)}]\cos \lambda \sin \lambda \\
&&+\frac 14(\widehat{J}_{+}^2e^{i2\varphi (t)}+\widehat{J}_{+}\widehat{J}%
_{-}+\widehat{J}_{-}\widehat{J}_{+}+\widehat{J}_{-}^2e^{-i2\varphi (t)})\sin
^2\lambda  \nonumber
\end{eqnarray}
we obtain apart from a trivial constant 
\begin{equation}
\widehat{H}^{\prime }=-G\widehat{J}_z-\frac q2\widehat{J}_z^2
\end{equation}
where the two-photon transition terms (proportional to $\widehat{J}%
_{+}^2e^{i2\varphi (t)}$ and $\widehat{J}_{-}^2e^{-i2\varphi (t)}$ ) have
been neglected as a reasonable approximation\cite{13,14} which is good
enough for the small $q$ in comparing with the transition coupling between
two components, namely, $q\ll G$. The Berry phase in this case is the same
as eq.(43) due to $\omega _0=0$. The interatom collisions do not affect the
Berry phase. This observation for the Berry phase is new at least for the
model we considered. The time-evolution operator is 
\begin{equation}
\widehat{U}_q(-\frac \pi 2,t)=\widehat{R}^{+}(-\frac \pi 2,t)e^{it(G\widehat{%
J}_z+\frac q2\widehat{J}_z^2)}\widehat{R}(-\frac \pi 2,0)
\end{equation}
The population imbalance for the initial state $|\psi _1(0)\rangle
=|j,j\rangle $ is able to be evaluated with the help of eq.(16) and the
Dicke-state representation of the SU(2) coherent state eq.(19) as 
\begin{equation}
\Delta N_1(q,t)=(\frac 12)^N\sum_{m=-\frac N2}^{\frac N2-1}(-1)^{N-2m}%
{N \choose \frac N2+m+1}%
(\frac N2+m+1)^{\frac 12}\cos \{[G+q(m+\frac 12)]t-\varphi (t)\}
\end{equation}
To compare the time-evolution of population imbalance obtained here with
that in the case of $q=0$ i.e. eq.(44) (for $m=j=\frac N2$ ) where the time
variation of imbalance is a simple oscillation, the time-evolution of
eq.(50) is shown in Fig.1 with various values of the ratio $\frac qG$ and
the number of particles $N$. It is seen that the damping goes faster when
the number of particles, $N$ , and the ratio $\frac qG$ increase (Fig.1
(a),(b)). The $N$-dependence of the oscillation of population imbalance for
fixed ratio $\frac qG$ is shown in Fig.1(a) for $N=10^2$ (dash line), $%
N=10^3 $ (dotted line), and $N=10^4$ (solid line) respectively. The
nonlinear interaction dependence for fixed number of particles, $N$ $=10^3$,
is shown in Fig.1(b) for $\frac qG=0.01\sim 0.1$. With the similar method
the time-evolution of population imbalance from the initial spin coherent
state eq.(16), $|\psi _2(0)\rangle =|{\bf n}\rangle $, with nonvanishing $q$
is obtained as 
\begin{eqnarray}
\Delta N_2(q,t) &=&-\sum_{m=-\frac N2}^{\frac N2-1}%
{N \choose \frac N2+m+1}%
(\frac N2+m+1)^{\frac 12}\cos ^{N+2m+1}(\frac{-\frac \pi 2+\theta }2)\sin
^{N-2m-1}(\frac{-\frac \pi 2+\theta }2)  \nonumber \\
&&\times \cos \{[G+q(m+\frac 12)]t-\phi -\varphi (t)\}
\end{eqnarray}
where we again use the Dicke-state representation of SU(2) coherent state
eq.(19). The oscillation of population imbalance is shown in Fig.2 with
various values of the coupling strength $\frac qG$ and angle $\theta $ (for
the sake of simplicity we set $\phi =\varphi (t)=0$). Besides the damping
the most important effect of the nonlinear interaction with initial spin
coherent state is the collapse-revival of the population- imbalance
oscillation. The particle number dependence of collapse-revival is shown in
Fig.2 (a) for fixed $\frac qG$ and the parameter $\theta $. We observe an
interesting phenomena that the frequency of the collapse-revival depends on
both the number of particles and the coupling strength $\frac qG$. In Fig.2
(b,c) we show the frequency behavior of the collapse-revival varying with
the product of particle number and coupling strength , $N\frac qG$. The
frequency is almost the same for the same value of the product $N\frac qG$
while different individual values of $N$ and $\frac qG$. Moreover $N\frac qG$%
-dependence of collapse-revival frequency is not monotonic. A critical value 
$N\frac qG=65$ is found at which the frequency of collapse-revival
approaches a minimum. To see the effect of nonlinear interatom collisions
closely we look at the population imbalance eq.(39) with $\phi =\varphi
(t)=0 $ for the case of $q=0$ as a comparison. In that case the population
imbalance of eq.(39) reduces to a simple oscillation such that 
\begin{equation}
\Delta N_2(t,\omega _0=0)=\frac N2\cos \theta \cos (2Gt)
\end{equation}
for $\omega _0=0$. It is obviously that the nonlinear interaction results in
both damping and collapse-revival of population-imbalance oscillation. The
simple oscillation of eq.(52) with Rabi frequency is in agreement with the
experimental observation \cite{2}.

Particularly the initial state can be prepared such that $\theta =\frac \pi 2
$, the population imbalance vanishes for both the cases with and without the
nonlinear interatom collisions (see eq.(52)). While the phase locking state
is achieved again by seeing that 
\begin{equation}
\langle \widehat{J}_{+}(q,t)\rangle _2(\phi =0,\theta =\frac \pi 2)=\frac N2%
e^{-i\varphi (t)}
\end{equation}
which is the same as eq.(46) for the case of $q=0$. In other words the phase
locking is independent of the nonlinear interaction. The expectation value, $%
\langle \widehat{J}_{+}(q,t)\rangle _2$, for general $\phi $, $\theta $ is
also derived analytically with the help of eqs. (16),(19). The resulting
formula is tedious and may not be of interest to be presented here.

\section{Conclusion}

Using the exact solution of time-dependent Schr$\ddot{o}$dinger equation the
population imbalance and phase dynamics are evaluated with various initial
states particularly with SU(2) coherent states which are the most realistic
initial states for the two-species BEC created by coupling two hyperfine
states of \thinspace atoms with radiation field\cite{12}. We conclude that
the self-trapping can be achieved from initial SU(2) coherent state with
asymmetric trap potential only. The phase locking is obtained also from the
initial SU(2) coherent state and is independent of the nonlinear
interaction, which may be observed experimentally in terms of Ramsey's
method measuring the relative phase of two components of BEC in different
hyperfine atomic states. The nonlinear interatom coupling does not affect
the Berry phase but leads to the damping and collapse-revival of the
population imbalance oscillations. The interesting particle number
dependence of the collapse-revival and Berry phase as well is explored.

{\bf Acknowledgment:} This work was supported by the Natural Science
Foundation of China under Grant Nos. 10075032.

\bigskip

\bigskip

{\bf Appendix:}

\bigskip It is easy to prove the following useful relations\cite{20}, 
\begin{equation}
\widehat{R}\widehat{J}_{+}\widehat{R}^{+}=\widehat{J}_{+}\cos ^2\frac \lambda
2-\widehat{J}_{-}e^{-2i\varphi (t)}\sin ^2\frac \lambda 2-\widehat{J}%
_ze^{-i\varphi (t)}\sin \lambda  \eqnum{A1}
\end{equation}
\begin{equation}
\widehat{R}\widehat{J}_{-}\widehat{R}^{+}=\widehat{J}_{-}\cos ^2\frac \lambda
2-\widehat{J}_{+}e^{2i\varphi (t)}\sin ^2\frac \lambda 2-\widehat{J}%
_ze^{i\varphi (t)}\sin \lambda  \eqnum{A2}
\end{equation}
\begin{equation}
\widehat{R}\widehat{J}_z\widehat{R}^{+}=\widehat{J}_z\cos \lambda +\frac 12(%
\widehat{J}_{+}e^{i\varphi (t)}+\widehat{J}_{-}e^{-i\varphi (t)})\sin \lambda
\eqnum{A3}
\end{equation}
\begin{equation}
i\widehat{R}\frac \partial {\partial t}\widehat{R}^{+}=2\frac{d\varphi }{dt}%
\sin ^2\frac \lambda 2\widehat{J}_z-\frac 12\frac{d\varphi }{dt}\sin \lambda
(e^{i\varphi }\widehat{J}_{+}+e^{-i\varphi }\widehat{J}_{-})+\frac i2\frac{%
d\lambda }{dt}(e^{i\varphi }\widehat{J}_{+}-e^{-i\varphi }\widehat{J}_{-}). 
\eqnum{A4}
\end{equation}

The time-dependent angular momentum operators $\widehat{J}_z$ and $\widehat{J%
}_{+}$ are obtained respectively as

\bigskip

\bigskip 
\begin{eqnarray}
\widehat{J}_z(t) &=&\widehat{U}^{+}(t)\widehat{J}_z\widehat{U}(t)=[\cos
^2\lambda +\sin ^2\lambda \cos (\alpha t+\varphi (t))]\widehat{J}_z 
\eqnum{A5} \\
&&+\frac 12\sin \lambda [-\cos \lambda +\cos ^2\frac \lambda 2e^{-i(\alpha
t+\varphi (t))}-\sin ^2\frac \lambda 2e^{i(\alpha t+\varphi (t))}]\widehat{J}%
_{+}  \nonumber \\
&&+\frac 12\sin \lambda [-\cos \lambda +\cos ^2\frac \lambda 2e^{i(\alpha
t+\varphi (t))}-\sin ^2\frac \lambda 2e^{-i(\alpha t+\varphi (t))}]\widehat{J%
}_{-},  \nonumber
\end{eqnarray}

\bigskip

\bigskip and

\bigskip \ 
\begin{eqnarray}
\widehat{J}_{+}(t) &=&\sin \lambda [\cos ^2\frac \lambda 2e^{i\alpha
(t)}-\sin ^2\frac \lambda 2e^{-i[\alpha (t)+2\varphi (t)]}-\cos \lambda
e^{-i\varphi (t)}]\widehat{J}_z  \eqnum{A6} \\
&&+[\cos ^4\frac \lambda 2e^{i\alpha (t)}+\sin ^4\frac \lambda 2e^{-i[\alpha
(t)+2\varphi (t)]}+\frac 12\sin ^2\lambda e^{-i\varphi (t)}]\widehat{J}_{+} 
\nonumber \\
&&+\{-\cos ^2\frac \lambda 2\sin ^2\frac \lambda 2[e^{i\alpha
(t)}+e^{-i[\alpha (t)+2\varphi (t)]}]+\frac 12\sin ^2\lambda e^{-i\varphi
(t)}\}\widehat{J}_{-}.  \nonumber
\end{eqnarray}

\bigskip

\bigskip

Figure Captions

Fig.1: Time-evolution of population imbalance from initial state $|\psi
_1(0)\rangle =|j,j\rangle $, $\varphi (t)=0$. (a)$\frac qG=0.01$; $N=10^2$, $%
10^3$, $10^4$. (b)$N=10^3$, $\frac qG=0.01\sim 0.1$.

Fig.2: Time-evolution of population imbalance from initial state $|\psi
_2(0)\rangle =|{\bf n}\rangle $, $\varphi (t)=\phi =0$, $\theta =0.8\frac \pi
2$ . (a)$\frac qG=0.01$, $N=10^3$ $\sim 8\times 10^3$. (b) $N\frac qG$
dependence of $\Delta N_2(q,t)$, for $N\frac qG=60\sim 70$, Left $N=2\times
10^3$; Right $N=3\times 10^3$. (c)$N\frac qG$ dependence of $\Delta N_2(q,t)$%
, for $N\frac qG=60\sim 70$, Left $N=4\times 10^3$; Right $N=5\times 10^3$.


\bigskip

\bigskip

\bigskip

\bigskip

\begin{references}
\bibitem{1}  M. R. Andrews, C. G. Townsend, H.-J. Miesner, D. S. Durfee, D.
M. Kurn, and W. Ketterle, Science, {\bf 275}, 637 (1997)

\bibitem{2}  D. S. Hall, \ M. R. Matthews, J. R. Ensher, C. E. Wieman, and
E. A. Cornell, Phys. Rev. Lett. {\bf 81}, 1543 (1998)

\bibitem{3}  D. S. Hall, {\it Bose-Einstein Condensates and Atom Lasers,}
Edited by S. Martellucci, A. N. Chester, A. Aspect, and M. Inguscio, (Kluwer
Academic/Plenum Publishers 2000), PP 31.

\bibitem{4}  G. J. Milburn, J. Corney, E. M. Wright, and D. F. Walls, Phys.
Rev. A {\bf 55}, 4318 (1997).

\bibitem{5}  A. Smerzi, S. Fantoni, S. Giovanazzi, and S. R. Shenoy, Phys.
Rev. Lett. {\bf 79}, 4950 (1997).

\bibitem{6}  S. Raghavan, A. Smerzi, S. Fantoni, and S. R. Shenoy, Phys.
Rev. A {\bf 59}, 620 (1999).

\bibitem{7}  J. Williams, R. Walser, J. Cooper, E. Cornell, and M. Holland,
Phys. Rev. A {\bf 59}, R31 (1999).

\bibitem{8}  S. Raghavan, A. Smerzi, and V. M. Kenkre, Phys. Rev. A {\bf 60}%
, R1787 (1999).

\bibitem{9}  J. Javanainen and M. Y. Ivanov, Phys. Rev. A {\bf 60, }2351
(1999).

\bibitem{10}  M. J. Steel and M. J. Collett, Phys. Rev. A {\bf 57}, 2920
(1998).

\bibitem{11}  A. Smerzi and S. Raghavan, Phys. Rev. A {\bf 61}, 063601
(2000).

\bibitem{12}  D. Gordon and C. M. Savage, Phys. Rev. A {\bf 59}, 4623 (1999).

\bibitem{13}  L.-M. Kuang and Z.-W. Ouyang, Phys. Rev. A {\bf 61}, 023604
(2000).

\bibitem{14}  W.D. Li, X.J. Zhou, Y.Q. Wang, J.Q. Liang, and W.M. Liu, Phys.
Rev. A{\bf 64}, 015602 (2001).

\bibitem{15}  N. Ramsey, {\it Molecular Beams} (Claredon Press, Oxford,
1956).

\bibitem{16}  I. Fuentes-Guridi, J Pachos, S. Bose, V. Vedral, and S. Choi,
Phys. Rev. A{\bf 66}, 022102 (2002)

\bibitem{17}  M. V. Berry, Proc. Roy. Soc. A {\bf 392}, 45 (1984)

\bibitem{18}  A. Perelomov, {\it Generalized Coherent States and Their
Applications} (Springer, Heidelberg, 1986).

\bibitem{19}  J.-Q. Liang and H.J.W. M$\ddot{u}$ller-Kirsten, Ann. Phys. 
{\bf 219}, 42 (1992)

\bibitem{20}  Y.-Z. Lai, J.-Q. Liang, H.J.W. M$\ddot{u}$ller-Kirsten, and
J.-G. Zhou, Phys. Rev. A{\bf 53}, 3691 (1996)

\bibitem{21}  A. Inomata, H. Kuratsuji, and C.C. Gerry, {\it Path Integrals
and Coherent States of SU(2) and SU(1,1),} (World Scientific, 1992)

\bibitem{22}  D. Loss, D.P. DiVincenzo, G. Grinstein, Phys. Rev. Lett. {\bf %
69}, 3232, (1992)

\bibitem{23}  J.-Q. Liang, H.J.W. M$\ddot{u}$ller-Kirsten, J.-G. Zhou, Z.
Phys. B {\bf 102}, 525, (1997)

\bibitem{24}  Z.-D. Chen, J.-Q. Liang, F.-C. Pu, Phys. Lett. A {\bf 300},
654, (2002)\quad
\end{references}
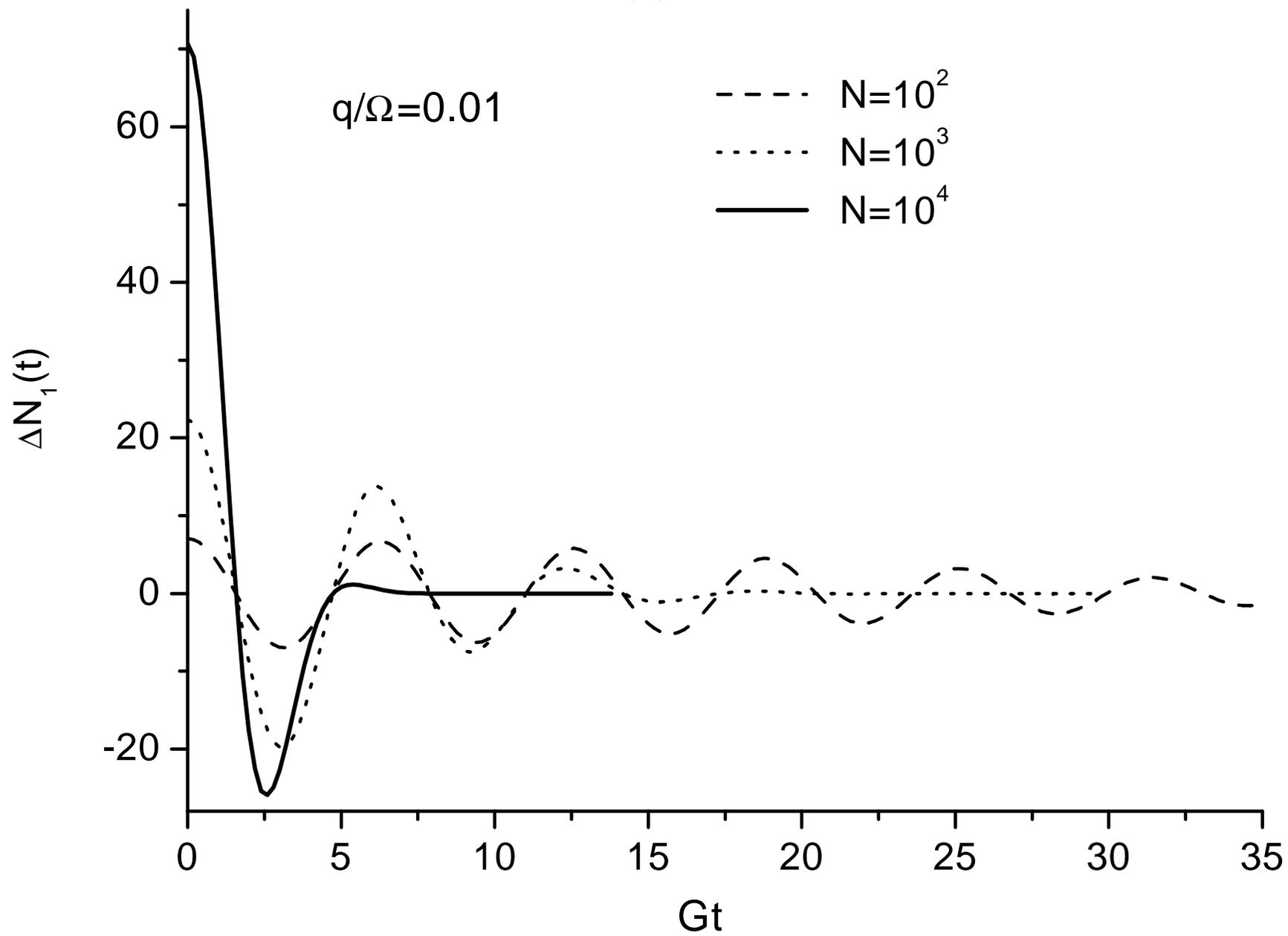
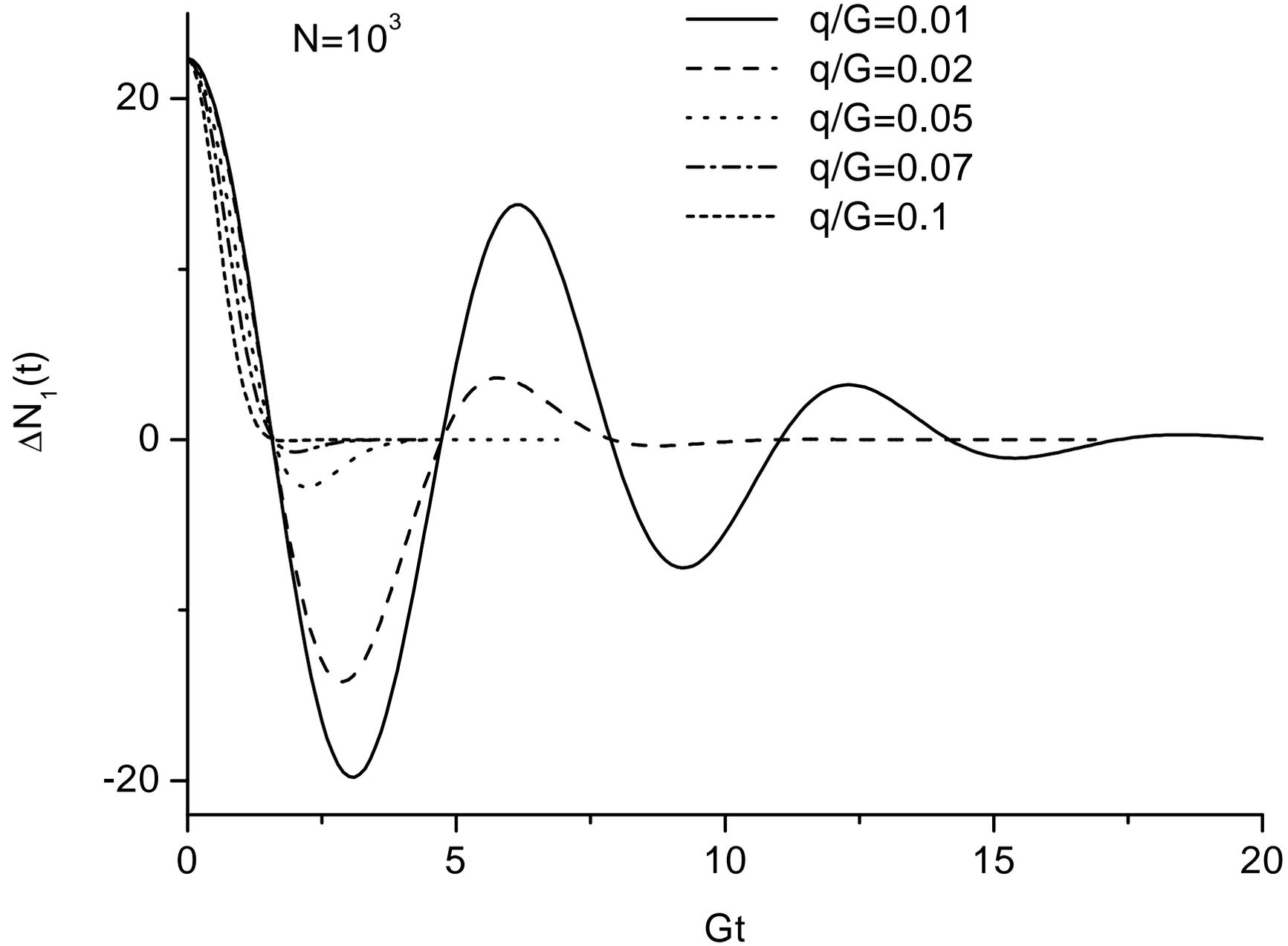
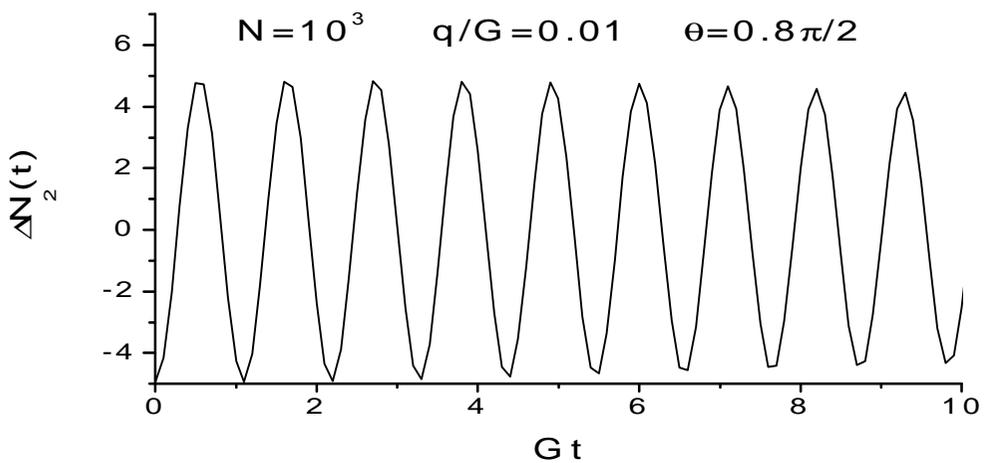
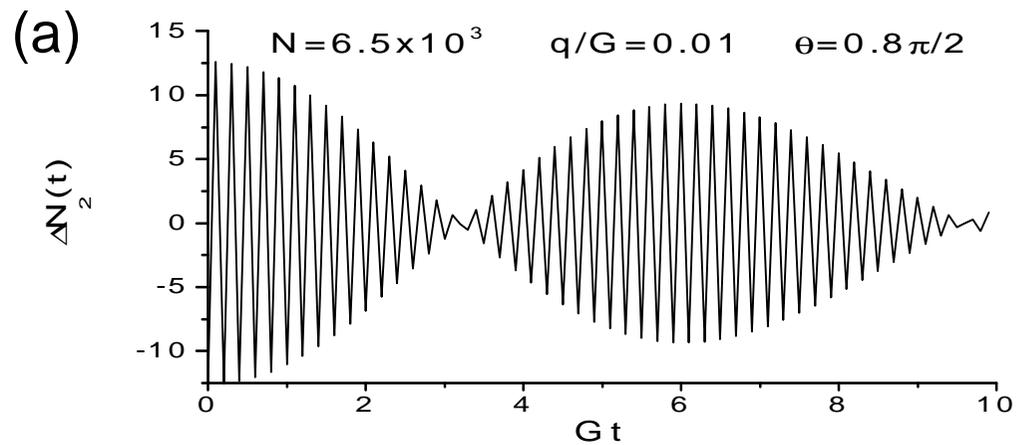
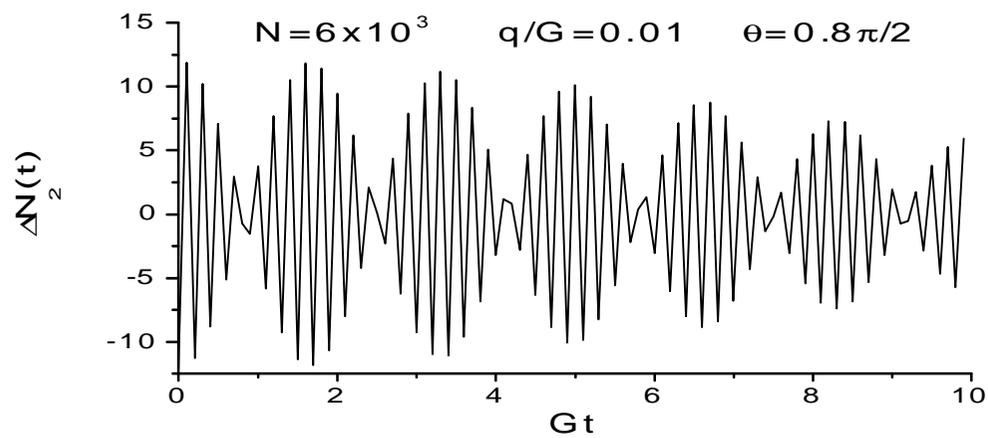
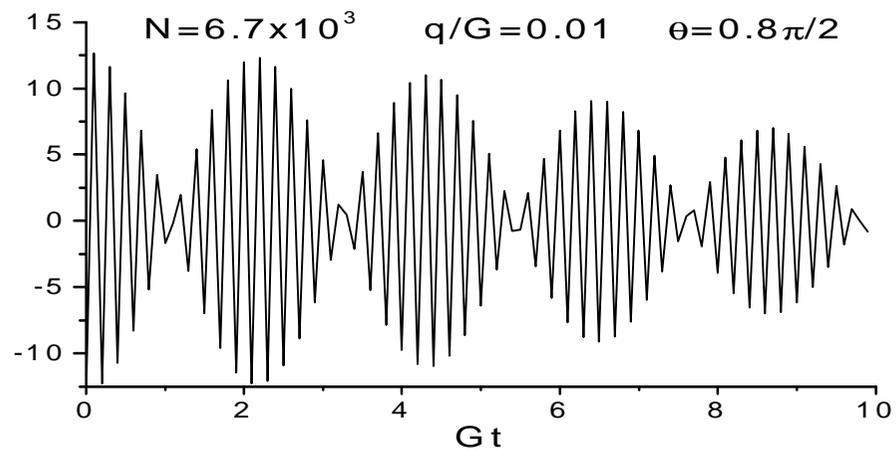
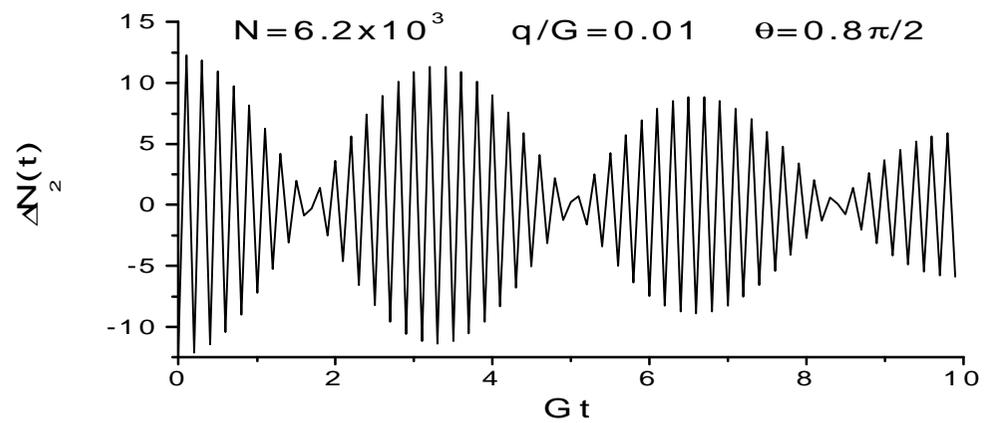
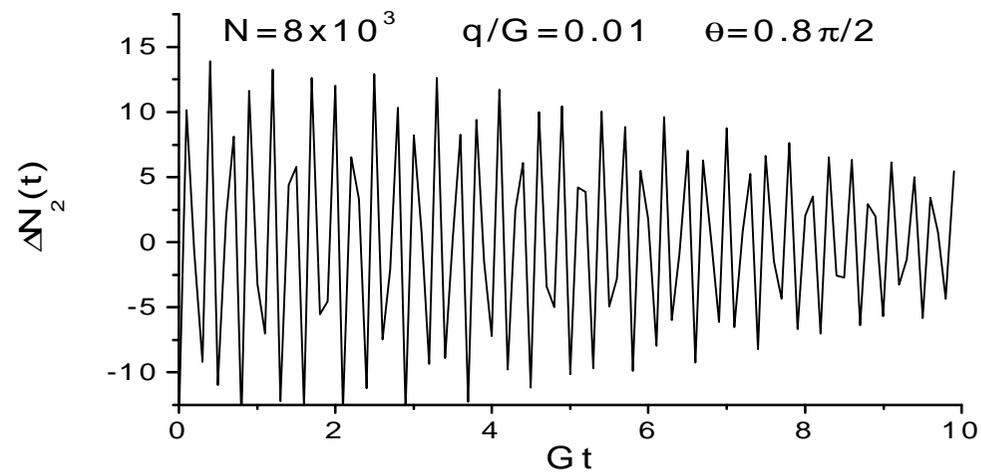
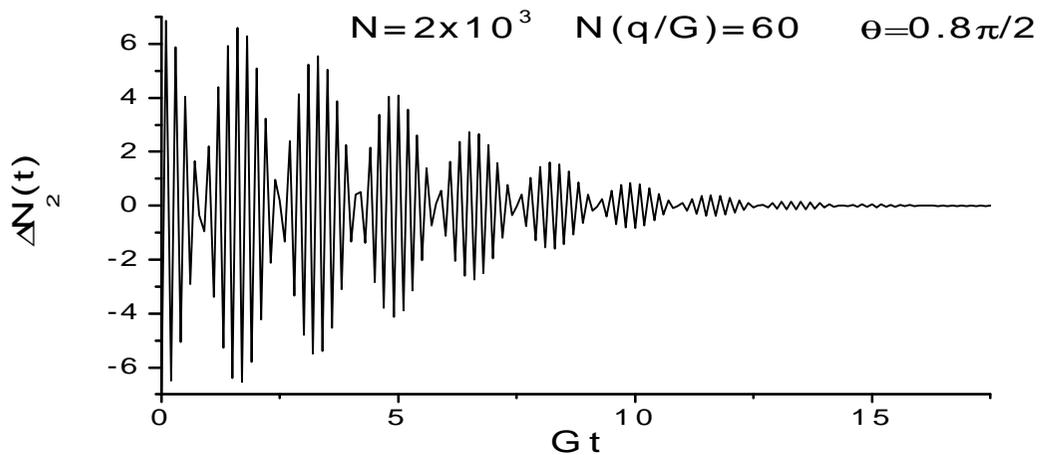
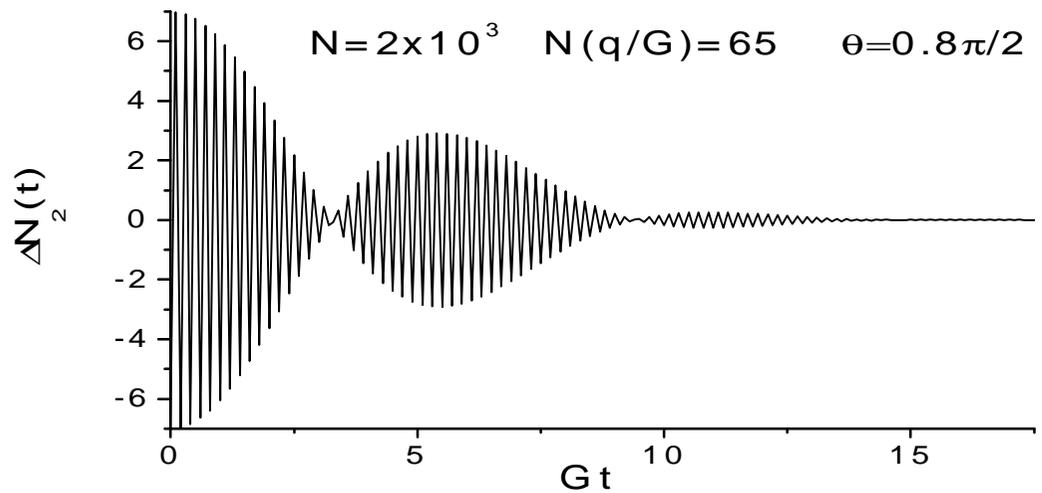
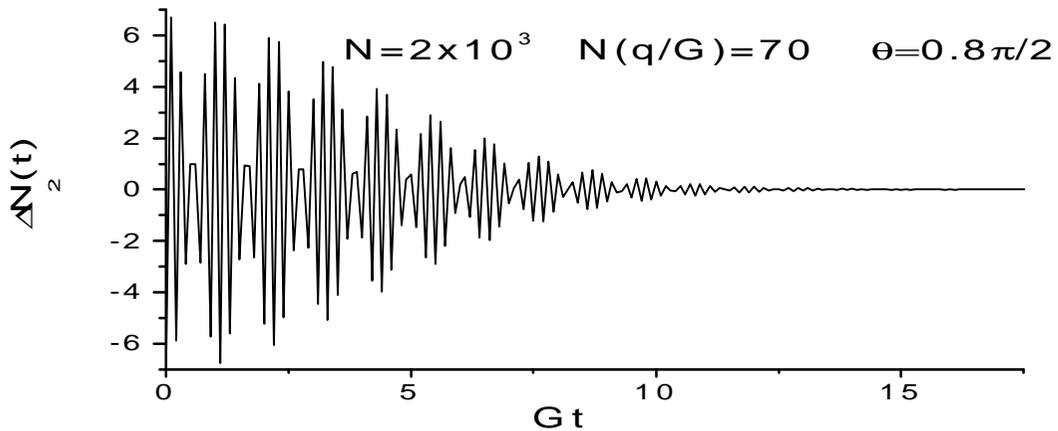
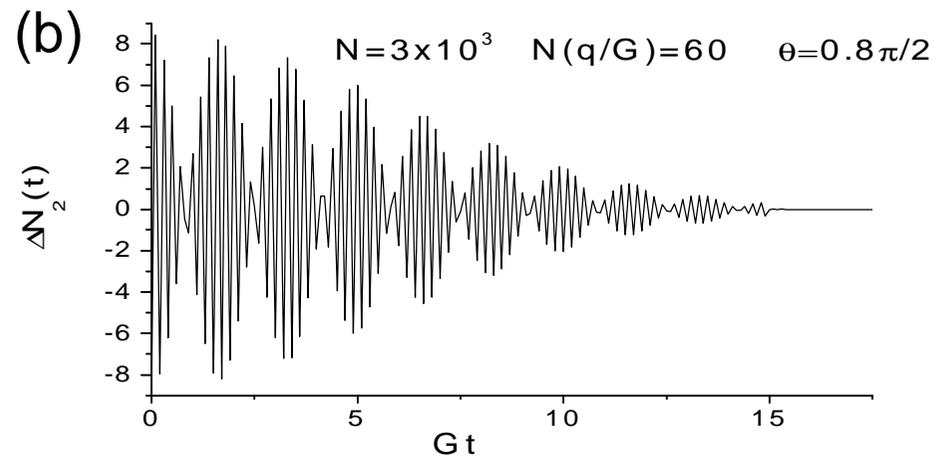
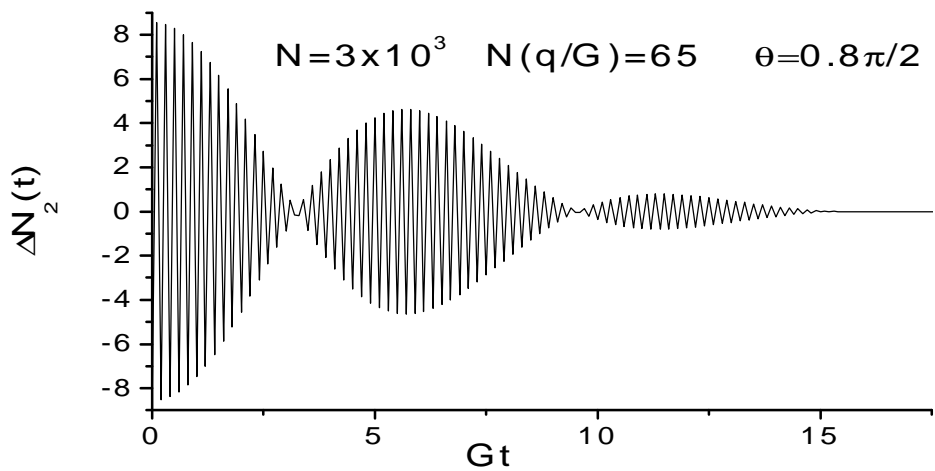
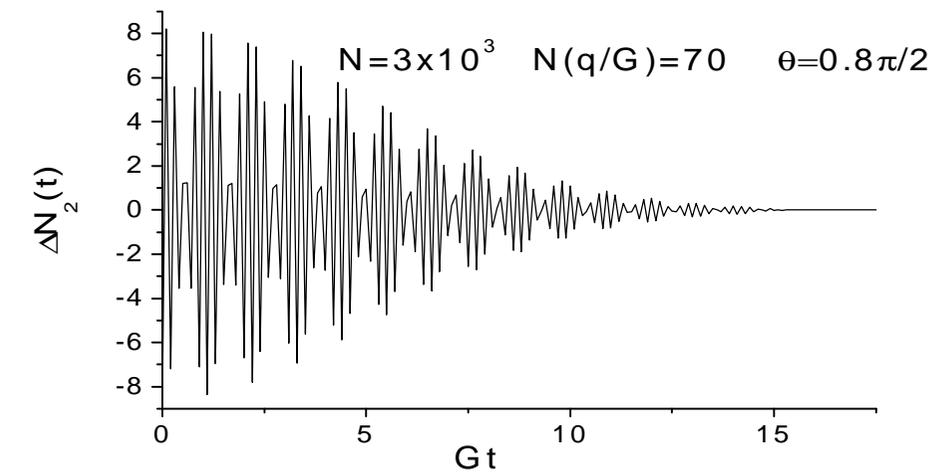
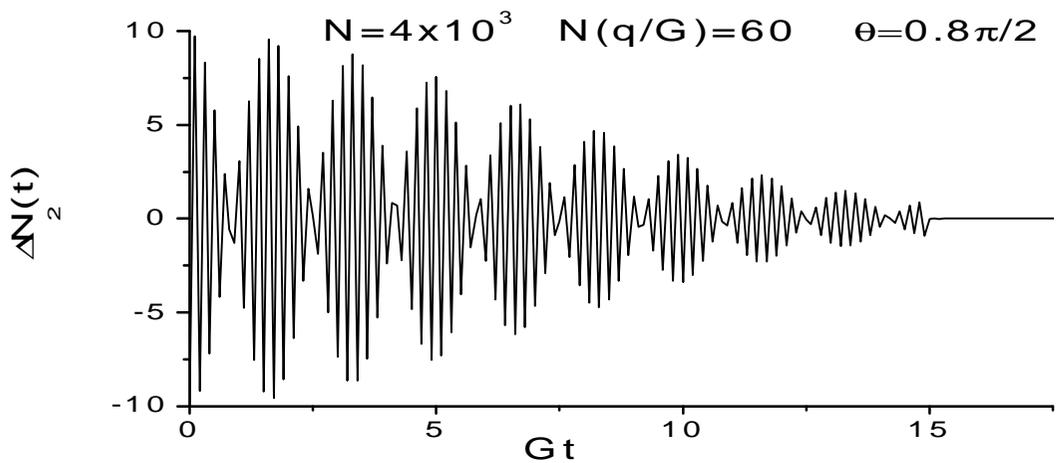
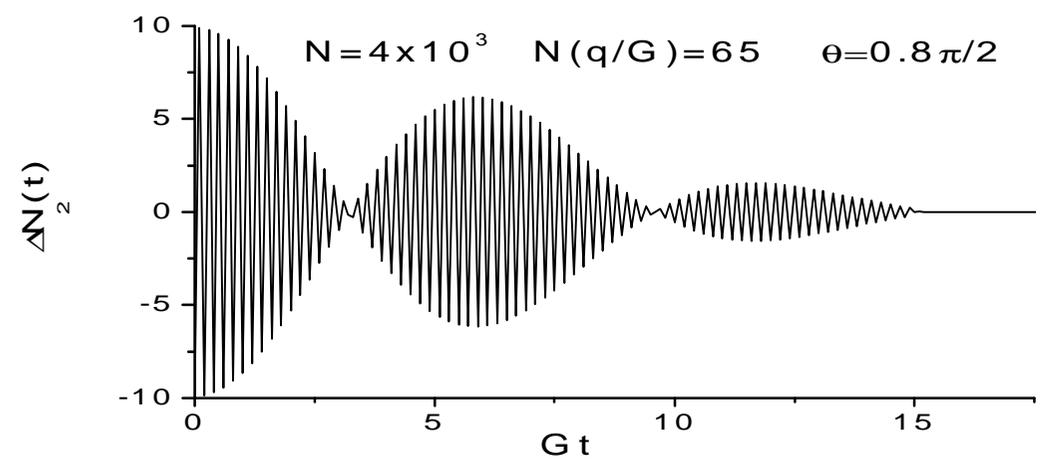
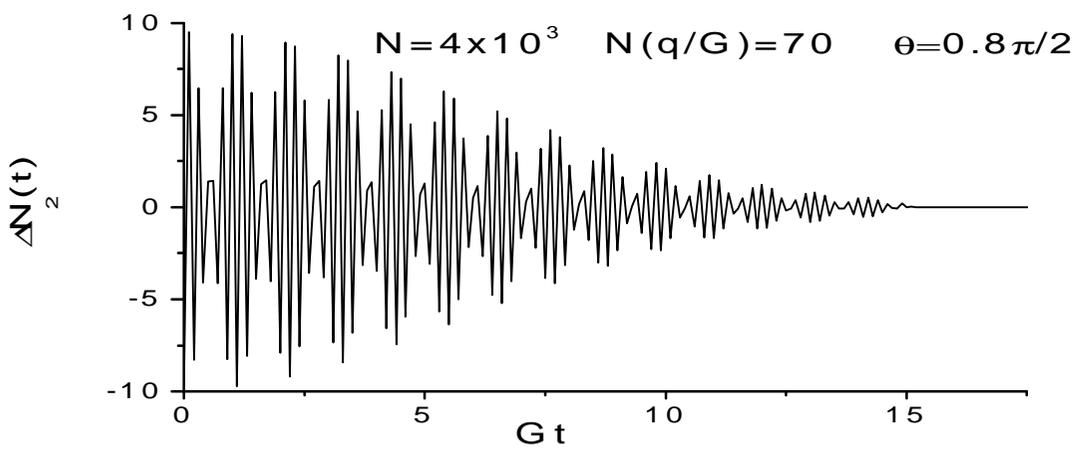
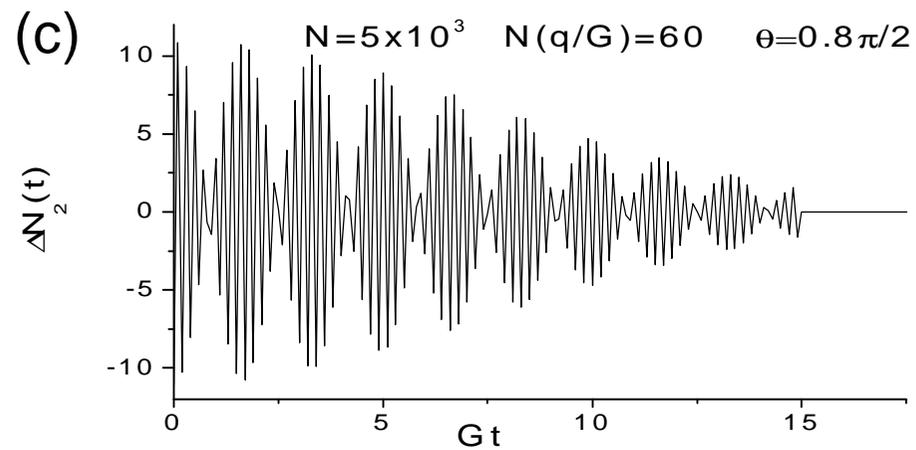
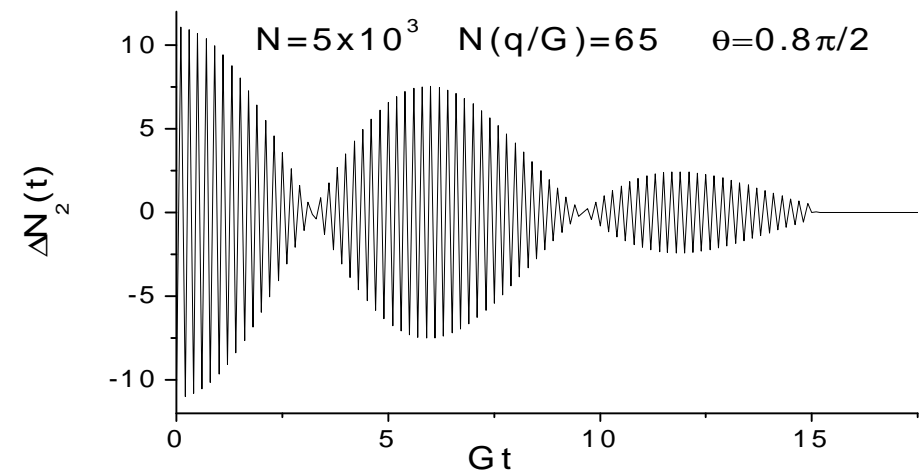
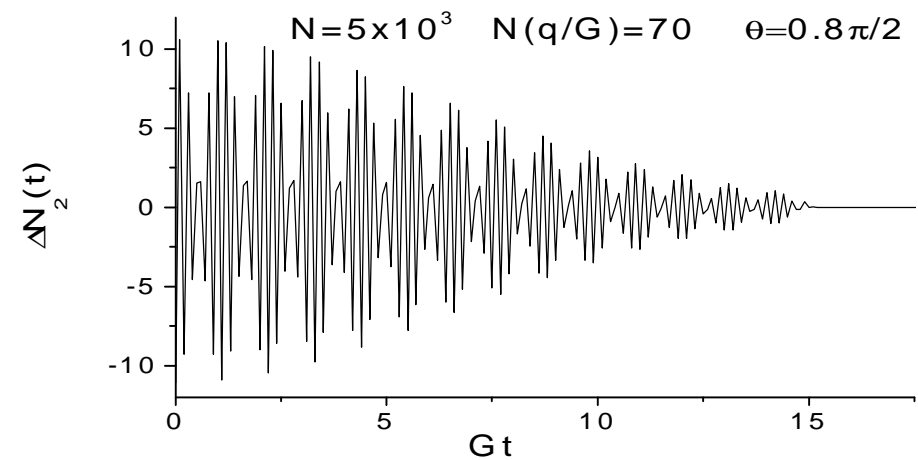
\end{document}